\documentclass{emulateapj}
\usepackage{natbib}
\usepackage{amsmath}
\usepackage{subfigure}
\graphicspath{{figs/}}

\newcommand{\pd}[2]{ \frac{\partial #1}{\partial #2} }

\newcommand{\vct}[1]{ \mathbf #1 }
\newcommand{\vsp}[1]{ \mathbf #1 }
\newcommand{\figext}[1]{#1.eps}

\makeatletter
\def\ScaleIfNeeded{%
  \ifdim\Gin@nat@width>\linewidth
  \linewidth
  \else
  \Gin@nat@width
  \fi
}
\makeatother

\shorttitle{Numerical Simulations of Relativistic MHD Turbulence}
\shortauthors{J. Zrake and A. I. MacFadyen}

\begin{document}

\title{Numerical Simulations of Driven Relativistic MHD Turbulence}
\author{Jonathan Zrake and Andrew I. MacFadyen}

\affil{Center for Cosmology and Particle Physics, Physics Department,
  New York University, New York, NY 10003, USA}


\begin{abstract}

  A wide variety of astrophysical phenomena involve the flow of
  turbulent magnetized gas with relativistic velocity or energy
  density. Examples include gamma-ray bursts, active galactic nuclei,
  pulsars, magnetars, micro-quasars, merging neutron stars, X-ray
  binaries, some supernovae, and the early universe. In order to
  elucidate the basic properties of the relativistic
  magnetohydrodynamical (RMHD) turbulence present in these systems, we
  present results from numerical simulations of fully developed driven
  turbulence in a relativistically warm, weakly magnetized and mildly
  compressible ideal fluid. We have evolved the RMHD equations for
  many dynamical times on a uniform grid with $1024^3$ zones using a
  high order Godunov code. We observe the growth of magnetic energy
  from a seed field through saturation at $\sim 1\%$ of the total
  fluid energy. We compute the power spectrum of velocity and
  density-weighted velocity $U = \rho^{1/3} v$ and conclude that the
  inertial scaling is consistent with a slope of $-5/3$. We compute
  the longitudinal and transverse velocity structure functions of
  order $p$ up to $11$, and discuss their possible deviation from the
  expected scaling for non-relativistic media. We also compute the
  scale-dependent distortion of coherent velocity structures with
  respect to the local magnetic field, finding a weaker scale
  dependence than is expected for incompressible non-relativistic
  flows with a strong mean field.

  \vspace{0.1in}
\end{abstract}

\section{Introduction}
Turbulence is a fundamental open problem in classical physics and is
of broad importance in science and technology. While non-relativistic
turbulence has been studied in great detail (e.g., Davidson 2004),
relatively little attention has been devoted to the properties of
turbulence in relativistic gas. Yet gas flows in astrophysics are
known to be relativistic and turbulent in a wide variety of systems
under active investigation. Extensive evidence of relativistic gas
flow now exists, most dramatically in astronomical observations of
outflows from gamma-ray bursts (GRBs) \citep[see recent reviews
by][]{Piran:2004p4605, Fox:2006p4649, Woosley:2006p4650,
  Zhang:2004p4732, Gehrels:2009p4728} and active galactic nuclei
\citep[e.g.][]{Begelman:1984p4745, Antonucci:1993p4756,
  Krolik:1999p4787, Frank:2002p4792}. Nearer by, relativistic outflows
are observed from a variety of sources including micro-quasars, soft
gamma repeaters (SGRs), and some supernovae
\citep{Soderberg:2010p4890}. Theoretical models for the central
engines powering a broad range of astrophysical phenomena invoke gas
accretion onto black holes and neutron stars. In addition, the
gravitational wave sources targeted by LIGO involve the hydrodynamical
merging of neutron star--neutron star or neutron star--black hole
binaries. The merging of neutron star binaries generates shear as the
stars touch and merge. Shear is also generated as neutron stars are
shred when merging with a black hole binary companion.
Kelvin-Helmholtz instability in these shear flows generate turbulence
which can lead to large amplifications of magnetic field
\citep{Zhang:2009p3418}. Such field amplification may be crucial for
creating conditions capable of extracting observed luminosities from
the GRB central engine \citep{Gehrels:2009p4728}.

Many astrophysical flows, including the aforementioned, are at least
partially relativistic (bulk and/or thermal Lorentz factor $\gtrsim
1$) and all are highly susceptible to turbulence due to the extremely
high Reynolds numbers characterizing astrophysical gas. Knowledge of
the properties of relativistic turbulence is also of general
importance in physics, with direct applications to early universe
cosmology, heavy ion colliders, and high energy density physics and
laboratory plasmas. Research elucidating the basic properties of
relativistic turbulence is thus broadly motivated.

In order to advance knowledge of turbulence in the largely unexplored
relativistic case, we have begun a series of numerical simulations of
turbulent flows in magnetized gas with relativistic energy density or
velocity. In this first in a series of papers on this topic we present
simulations and analysis of driven turbulence in a trans-sonic,
super-Alfv\'{e}nic medium for which the ratio of internal to rest mass
energy density is of order unity. Differences from non-relativistic
turbulence may be expected since pressure and magnetic fluctuations
communicated by acoustic and MHD waves modify the fluid inertial term
in the relativistic case.  This is distinct from the non-relativistic
case where fluid inertia is proportional to rest mass density but does
not depend on internal energy, pressure or magnetic field strength.

We have studied the same model (SATS1) at a range of resolutions up to
$1024^3$ in order to demonstrate numerical convergence. We make
comparisons with the extensive body of literature available for
compressible non-relativistic hydrodynamic and MHD turbulence
\citep[e.g.,][]{Padoan:1997p2749, VazquezSemadeni:2000p2760,
  Cho:2003p3040, Beresnyak:2005p4444, Kritsuk:2007p3858,
  Kritsuk:2009p2270, Schmidt:2008p4878, Federrath:2010p4876,
  Lemaster:2009p2711, Burkhart:2009p4462, Kowal:2010p3546}. We also
draw from studies of Alfv\'{e}nic turbulence in the non-relativistic
\citep[e.g.,][]{Politano:1995p4542, Cho:2000p4507, Maron:2001p3039,
  Beresnyak:2005p4444, Beresnyak:2009p4471, Beresnyak:2010p4470} and
relativistic force-free \citep{Cho:2005p4453} regimes. By applying the
same analyses utilized in these studies to our own models, our aim is
to elucidate the similarities and differences existing between the
relativistic and non-relativistic limit of MHD turbulence.

This paper is organized as follows: In \S \ref{sec:methods} we provide
the details of our computational scheme and problem setup. In \S
\ref{sec:results} we report on the saturation of magnetic energy and
provide power of spectra various quantities of interest. We also
investigate the scaling of various one and two dimensional structure
functions of velocity field. In \S \ref{sec:conclusions} we summarize
our findings.

\section{Methods}
\label{sec:methods}

\subsection{RMHD formulation}
We have employed Mara, a new unsplit, second-order Godunov code which
has been written to achieve robust and accurate evolution of the RMHD
equations on three-dimensional rectilinear grids. Mara solves the
system of ideal RMHD equations in conservation law form. The covariant
formulation for RMHD can be expressed as a system of coupled
conservation laws for particle number $N^\mu = \rho u^\mu$ and the
energy-momentum of the fluid denoted by $T^{\mu \nu} = \rho h^* u^\mu
u^\nu + p^* g^{\mu \nu} - b^\mu b^\nu$, where $\rho$ is the rest-mass
density of the fluid and $u$ is its four-velocity and we use units for
which the speed of light $c=1$ as follows,
\begin{subequations}\label{eqn:rmhd-system}
  \begin{align}
    \nabla_\nu N^\nu &= 0 \\
    \nabla_\nu T^{\mu \nu} &= 0 \\
    \pd{\vsp B}{t} &= \vsp \nabla \times (\vsp v \times \vsp B)
  \end{align}
\end{subequations}
Here, $b^\mu = F^\mu_{\ \nu} u^\nu$ is the magnetic field four-vector,
and $h^* = 1+e^*+p^*/\rho$ is the total specific enthalpy, where $p^*
= p_g + b^2/2$ is the total pressure, $p_g$ is the gas pressure and
$e^* = e_{th} + b^2/2\rho$ is the total specific energy density with
$e_{th}$ being the thermal part. In this study System
\ref{eqn:rmhd-system} is closed by the adiabatic equation of state,
$p_g = \rho e_{th} (\Gamma-1)$ with $\Gamma = 4/3$. Since our equation
of state is not isothermal are are those used in most previous studies
(exceptions include \cite{Porter:2002p4511}), thermalization of the
kinetic energy into internal energy results in a marginal decline of
the sonic Mach number throughout the run. This is discussed in more
detail in \S \ref{sec:startup}.

The fact that $\rho h^*$ and $p^*$ and $b^\mu$ directly effect all
aspects of the energy-momentum dynamics of the fluid through their
contributions to $T^{\mu \nu}$ means that RMHD turbulence dynamics
will contain mode couplings not present in the non-relativistic case.
This is because acoustic and RMHD waves will alter eddy dynamics via
their modifications to fluid inertia, as mentioned above.  In this
study, fluctuations in the thermal energy and pressure due to
compressive waves will dominate magnetic fluctuations in $T^{\mu \nu}$
since we consider the case with relativistic thermal pressure, $p/\rho
\sim 1$, but sub-dominant magnetic field. Alfv\'{e}nic RMHD turbulence
for which magnetic energy density is significant or dominant ($b^2/2
\sim \rho$) is also of interest and will form the basis for a future
study.

The ideal RMHD equations can also be expressed in flux conservative
form amenable to numerical solution as
\begin{equation}\label{eqn:diff-cons-law}
  \pd{\vct{U}}{t} + \sum_{j=1}^{3} \pd{\vct{F^j}}{x^j} = 0
\end{equation}
where the conserved quantities
\begin{equation}\label{eqn:conserved}
  \vct{U} = 
  \left( \begin{array}{c}
      D \\ \tau \\ \vsp{S} \\ \vsp{B}
    \end{array}
  \right) = 
  \left( \begin{array}{c}
      \rho W \\
      \rho h^* W^2 - p^* - (b^0)^2 - D \\
      \rho h^* W^2 \vsp{v} - b^0 \vsp{b} \\
      \vsp{B}
    \end{array} \right)
\end{equation}
represent the lab frame particle number, total energy (excluding rest
mass), momentum, and magnetic induction respectively, and the
corresponding fluxes are
\begin{equation}
  \vct{F^j} = 
  \left( \begin{array}{c}
      D v^j \\
      \tau v^j - b^0 B^j / W + p^* v^j \\
      \vsp{S} v^j - \vsp{b} B^j / W + p^* \hat{\vsp{x}}^j \\
      -(\delta^m_k \delta^j_l - \delta^j_k \delta^m_l)v^k B^l \hat{\vsp{x}}_m
    \end{array} \right).
\end{equation}
By introducing the volume averaged conserved quantities,
\begin{equation}
  \vct{U}_{i,j,k} = \frac{1}{\Delta V}
  \int_{x_{i-1/2}}^{x_{i+1/2}}\int_{y_{j-1/2}}^{y_{j+1/2}}\int_{z_{k-1/2}}^{z_{k+1/2}}
  \vct{U(\vsp{x},t)} dV
\end{equation}
and rewriting Equation \ref{eqn:diff-cons-law} using the divergence
theorem, we obtain volume averaged time derivatives in each zone,
\begin{eqnarray}\label{eqn:method-of-lines}
  \pd{}{t} \vct{U}_{i,j,k} =
  &-&\frac{1}{\Delta x}\left(\hat{\vct{F}}^1_{i+1/2,j,k} - \hat{\vct{F}}^1_{i-1/2,j,k}\right) \\ \nonumber
  &-&\frac{1}{\Delta y}\left(\hat{\vct{F}}^2_{i,j+1/2,k} - \hat{\vct{F}}^2_{i,j-1/2,k}\right) \\
  &-&\frac{1}{\Delta z}\left(\hat{\vct{F}}^3_{i,j,k+1/2} - \hat{\vct{F}}^3_{i,j,k-1/2}\right)    \nonumber
\end{eqnarray}
where $\hat{\vct{F}}^j$, are the fluxes of conserved quantities
evaluated at the interfaces between cell volumes. Equation
\ref{eqn:method-of-lines} would in principle complete the integration
scheme for advancing the solution in time. However, in practice an
expression which is higher order in time is desirable to maintain
accuracy. In this study, Mara has been configured to use a second
order unsplit MUSCL-Hancock type integration scheme similar to the one
described in \cite{Mignone:2006p2905}, and differs mostly in that
magnetic fields are volume instead of area-averaged. The scheme is
parameterized around an approximate Reimann solver for obtaining the
intercell fluxes $\hat{\vct{F}}^j$, and a reconstruction algorithm for
interpolating primitive quantities to the zone interfaces. For
completeness, the full scheme is described below.

\begin{enumerate}
\item Starting with $\vct{U}^n$, invert Equation \ref{eqn:conserved}
  to obtain $\vct{P}^n$. The primitive quantities are $\vsp{P} =
  (\rho, p_g, \vsp{v}, \vsp{B})^T$.

\item Compute the spatial derivatives $\pd{\vct{P}^n}{x^j}$ in each
  zone using the reconstruction algorithm, and obtain the interpolated
  primitive quantities at the zone interfaces as follows:
  \begin{eqnarray*}
    \vct{P}^{j,n}_R &=& \vct{P}^n + \pd{\vct{P}^n}{x^j}\frac{\Delta x^j}{2} \\
    \vct{P}^{j,n}_L &=& \vct{P}^n - \pd{\vct{P}^n}{x^j}\frac{\Delta x^j}{2}
  \end{eqnarray*}

\item Obtain the conserved quantities $\vct{U}^{j,n+1/2}_L$ and
  $\vct{U}^{j,n+1/2}_R$ at the half time-step by applying the
  approximate Riemann solver for the transverse fluxes
  $\hat{\vct{F}}^{2,3}$, and a Hancock operator for the normal fluxes
  $\vct{F}^1$. There is one predicted conserved state
  $\vct{U}^{j,n+1/2}$ for each direction. For example, in the
  $x$-direction,
  \begin{eqnarray*}
    &&\vct{U}^{1,n+1/2}_{i,j,k} = \vct{U}^n_{i,j,k}
    -\frac{\Delta t}{2\Delta x}\left[\vct{F}^1(\vct{P^{1,n}}_{R,i,j,k}) -
      \vct{F}^1(\vct{P^{1,n}}_{L,i,j,k})\right] \\
    &-&\frac{\Delta t}{2\Delta y}\left[\hat{\vct{F}}^2(\vct{P^{2,n}}_{R,i,j,k},\vct{P^{2,n}}_{L,i,j+1,k}) -
      \hat{\vct{F}}^2(\vct{P^{2,n}}_{R,i,j+1,k},\vct{P^{2,n}}_{L,i,j,k})\right] \\
    &-&\frac{\Delta t}{2\Delta z}\left[\hat{\vct{F}}^3(\vct{P^{3,n}}_{R,i,j,k},\vct{P^{3,n}}_{L,i,j,k+1}) -
      \hat{\vct{F}}^3(\vct{P^{3,n}}_{R,i,j,k+1},\vct{P^{3,n}}_{L,i,j,k})\right]
  \end{eqnarray*}

\item Invert Equation \ref{eqn:conserved} to obtain
  $\vct{P}^{j,n+1/2}$ from $\vct{U}^{j,n+1/2}$ for each $j=1,2,3$.

\item Obtain the interpolated primitive quantities at the zone
  interfaces for the half time step. The spatial gradients are not
  recomputed, but are reused from the beginning of the time step.
  \begin{eqnarray*}
    \vct{P}^{j,n+1/2}_R &=& \vct{P}^{n+1/2} + \pd{\vct{P}^n}{x^j}\frac{\Delta x^j}{2} \\
    \vct{P}^{j,n+1/2}_L &=& \vct{P}^{n+1/2} - \pd{\vct{P}^n}{x^j}\frac{\Delta x^j}{2}
  \end{eqnarray*}

\item Complete the time integration by applying the fluxes obtained
  from the Riemann solver in each direction.
  \begin{eqnarray*}
    &&\vct{U}^{n+1}_{i,j,k} = \vct{U}^n_{i,j,k} \\
    &-&\frac{\Delta t}{\Delta x}\left[\hat{\vct{F}}^1(\vct{P^{1,n+1/2}}_{R,i,j,k},\vct{P^{1,n+1/2}}_{L,i+1,j,k}) -
      \hat{\vct{F}}^1(\vct{P^{1,n+1/2}}_{R,i+1,j,k},\vct{P^{1,n+1/2}}_{L,i,j,k})\right] \\
    &-&\frac{\Delta t}{\Delta y}\left[\hat{\vct{F}}^2(\vct{P^{2,n+1/2}}_{R,i,j,k},\vct{P^{2,n+1/2}}_{L,i,j+1,k}) -
      \hat{\vct{F}}^2(\vct{P^{2,n+1/2}}_{R,i,j+1,k},\vct{P^{2,n+1/2}}_{L,i,j,k})\right] \\
    &-&\frac{\Delta t}{\Delta z}\left[\hat{\vct{F}}^3(\vct{P^{3,n+1/2}}_{R,i,j,k},\vct{P^{3,n+1/2}}_{L,i,j,k+1}) -
      \hat{\vct{F}}^3(\vct{P^{3,n+1/2}}_{R,i,j,k+1},\vct{P^{3,n+1/2}}_{L,i,j,k})\right]
  \end{eqnarray*}

\end{enumerate}

\subsection{Riemann Solvers}\label{sec:riemann}
\begin{figure*}
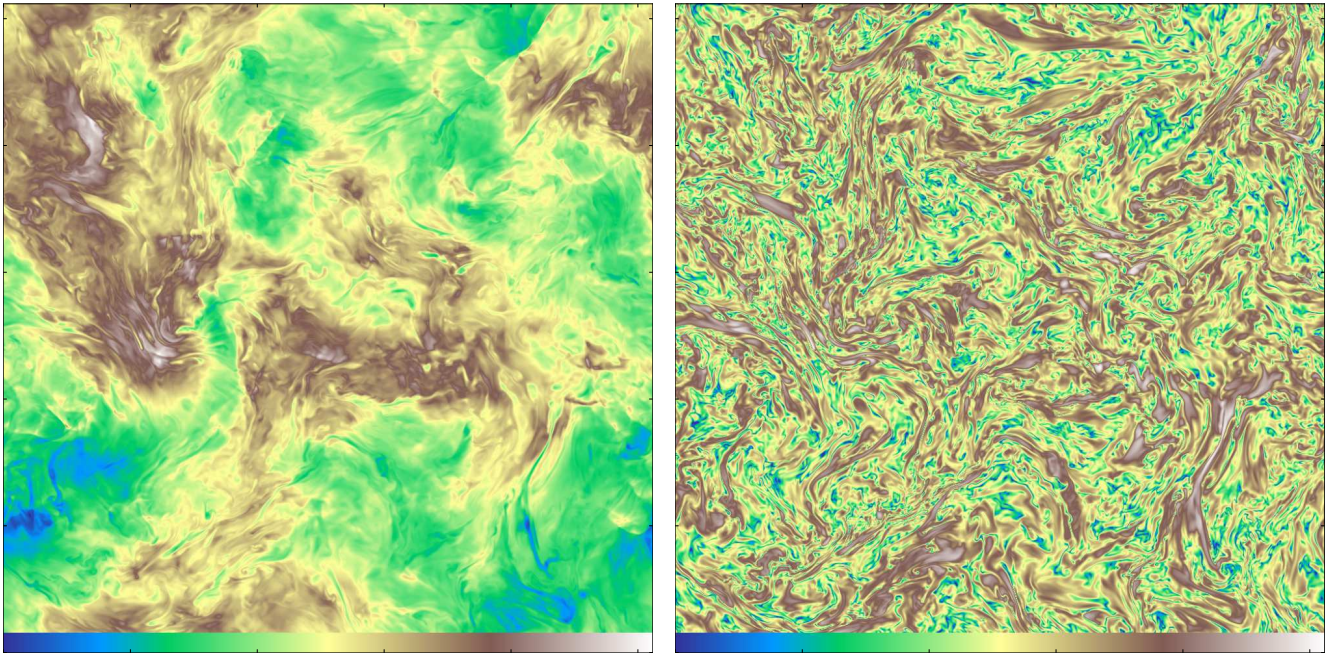

  \centering
  \subfigure{ \label{fig:rho}
    \includegraphics[width=3.4in]{\figext{SuperAlf-1024-t12-rho}}
  }
  \subfigure{ \label{fig:magP}
    \includegraphics[width=3.4in]{\figext{SuperAlf-1024-t12-magP}}
  }
  \caption{Slices of the domain taken at $t=12.0 T_{lc}$, normal to
    the $y$-axis. Shown is the rest-mass density (\emph{left}) scaled
    linearly between $0.326$ and $1.53$, and the ($\log_{10}$ of)
    magnetic pressure (\emph{right}), scaled between $3.16 \times
    10^{-4}$ and $1.26$.}
  \hspace{0.04\textwidth}
\end{figure*}
The fluid state in neighboring volumes is formally interpreted as the
initial data of a Riemann problem, the solution for which is
frequently sought by means of an exact or approximate Riemann solver
and yields the intercell fluxes $\hat{\vct{F}}^j$. In this study, we
have utilized the HLLD approximate Riemann solver
\citep{Mignone:2009p2269}, which computes Godunov fluxes at zone
interfaces by resolving the fast, Alfv\'{e}n, and contact waves. The
inclusion of these waves in the Godunov flux approximation has been
shown to be important for capturing the correct concentration of
magnetic energy, relative to the more diffusive HLLE and HLLC
approximate solvers \citep{Beckwith:2011p4448}.

Due to the fact that Mara employs ILES, the (effectively numerical)
viscous and resistive scales both occur near the grid spacing and thus
the magnetic Prandtl number $Pm=Rm/Re \sim 1$. However, deviations in
$Pm$ still occur based on the numerically dissipative behavior of the
scheme. Based on qualitative observations, the HLLC solver generates
roughly the same amount of numerical diffusion to the density and
pressure fields as HLLD, but substantially more to the magnetic
field. This is interpreted as the HLLD solver achieving a larger
numerical $Pm$. It is also important to note that as the numerical
dissipation goes down, the chances of encountering robustness issues
(see \S \ref{sec:robustness}) goes up.


\subsection{Magnetic field constraint}\label{sec:constrained-transport}
In this study, Mara uses volume-averaged magnetic fields which are
stored at cell centers. The solenoidal constraint $\nabla \cdot
\vsp{B} = 0$ (evaluated at cell corners) is maintained to machine
precision using the constrained transport method of
\cite{Toth:2000p2162}.

\subsection{Robustness}\label{sec:robustness}
Due to the complex nature of the RMHD equations, robustness concerns
are of great importance. In particular, RMHD, and GRMHD codes are
known to suffer from failures when inverting Equation
\ref{eqn:conserved} in order to obtain the primitive variables from
the conserved ones. This inversion can fail when the numerical root
finder (e.g. secant or Newton-Rapheson) fails to locate the root due
to an insufficiently close initial guess. But it can also fail when
the root corresponds to a state with negative pressure. Overcoming
these numerical limitations has been a major source of effort in
evolving the models used in this study.

The Mara code has been designed with an extensively tested and very
robust algorithm for the recovery of primitive variables. When the
failure is strictly numerical in nature, it will run through a series
of reset values and inversion relations. The default inversion
relation used in this study is adapted from \cite{Noble:2006p25} and
in the case of an adiabatic equation of state, may be solved using a
Newton-Rapheson iteration in the single unknown, $Z \equiv \rho h
W^2$.
\begin{eqnarray*}\label{eqn:noble-c2p}
  v^2      &=& \frac{S^2 Z^2 + (\vct{B} \cdot \vct{S})^2 (B^2 + 2Z)}{(B^2 + Z)^2 Z^2} \\
  \tau + D &=& \frac{B^2}{2}(1+v^2) - \frac{(\vct{B} \cdot \vct{S})^2}{2Z^2} + Z - p_g
\end{eqnarray*}
The starting value of $Z$ given to the root finder is evaluated from
the primitive variables at the previous time step. If a suitable
solution is not obtained, then the root finder is restarted using $Z =
\sqrt{D^2 + S^2}$, which becomes the exact solution in the limiting
case of weak magnetic field and small pressure. If a solution is still
not obtained, then the procedure is repeated using an inversion
relation adapted from \cite{Anton:2006p17}.
\begin{eqnarray*}\label{eqn:anton-c2p}
  S^2      &=& (Z + B^2)^2 \frac{W^2-1}{W^2} - (2Z + B^2)^2 \frac{(B^iS_i)^2}{Z^2} \\
  \tau + D &=& B + Z^2 - \frac{B^2}{2W^2} - \frac{(B^iS_i)^2}{Z^2} - p_g
\end{eqnarray*}
which can solved using a two-dimensional Newton-Rapheson algorithm for
the unknowns $Z$ and $W$. Because neither of these solvers contains
the other one's domain of success, there are circumstances where this
procedure obtains the solution even when the first solver or starting
value fails. However, there are states whose solution is not obtained
by any solver, or whose solution is unacceptable due to negative
pressure.

Under these conditions, the code assumes it has integrated the
conserved quantities into an unphysical configuration, and requires
further safety procedures in order not to crash. The safety feature we
have found to be the most practical is the addition of small amounts
of diffusion near ``unhealthy'' zones. This feature is implemented as a
Lax-Friedrichs flux, where the failed state $\vct{U}_{i,j,k} \mapsto
\vct{U'}_{i,j,k}$ according to
\begin{equation}\label{eqn:lax-diffusion}
  \vct{U'}_{i,j,k} = \vct{U}_{i,j,k} -
  \left(\tilde{\vct{f}}_{i+1/2,j,k} - \tilde{\vct{f}}_{i-1/2,j,k}\right) -
  \ldots
\end{equation}
where for brevity, we have written the fluxes in the $x$-direction
only, and
\begin{equation}
  \tilde{\vct{f}}_{i+1/2,j,k} = -\frac{r}{2d}
  (\vct{U}_{i+1,j,k} - \vct{U}_{i,j,k}) \theta_{i+1/2,j,k}
\end{equation}
$d=3$ is the number of dimensions, and $\theta_{i+1/2,j,k}=1$ if zones
$(i,j,k)$ or $(i+1,j,k)$ have been flagged as unhealthy, and $0$
otherwise. The effect of this prescription is to replace
$\vct{U}_{i,j,k}$ with a weighted average of itself and the average of
its neighboring cells, adding the most diffusion when $r \rightarrow
1$ and none when $r \rightarrow 0$. Throughout this study we have used
$r=0.2$. This formulation for the addition of diffusive terms has
several important features. Firstly, it naturally obeys the global
conservation of $\vct{U}$, but secondly it obeys the solenoidal
magnetic field constraint, because the constrained transport method
may be applied to the Lax-Friedrichs magnetic field fluxes before
adding them in Equation \ref{eqn:lax-diffusion}.

\subsection{Initial conditions and driving}\label{sec:initmodel-driving}
\begin{figure}
  \centering
  \includegraphics[width=\ScaleIfNeeded]{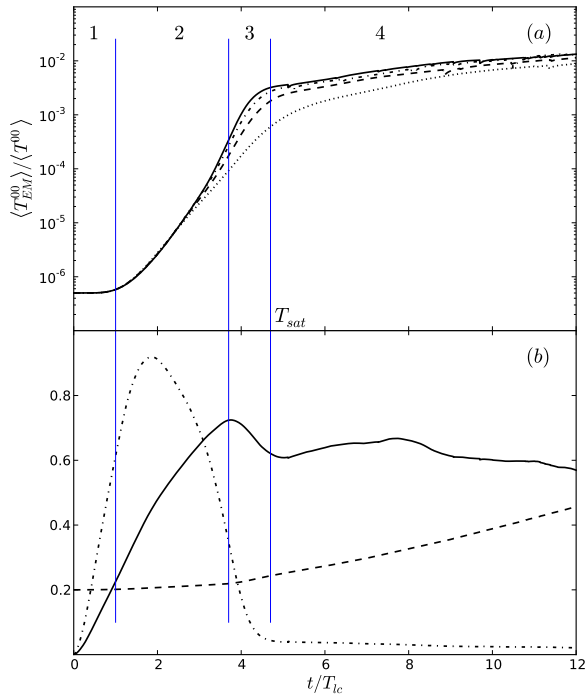}
  \caption{\emph{(a)} Shown are time histories indicating convergence
    of the magnetic energy fraction, the curves represent different
    resolutions: \emph{dotted}: $256^3$, \emph{dashed}: $512^3$,
    \emph{dash-dotted}: $768^3$, \emph{solid}: $1024^3$. \emph{(b)}
    Time histories of various quantities in our highest resolution
    model, $1024^3$. Shown are the volume-averaged sonic Mach number
    (\emph{solid}), Alfv\'{e}nic Mach number (\emph{dash-dotted}, divided
    by 200), and the internal energy (\emph{dashed}, divided by
    5). The Alfv\'{e}nic Mach number at $t=12T_{lc}$ is $\sim 4$.}
  \hspace{0.04\textwidth}
  \label{fig:time-hist}
\end{figure}

Our simulations take place in the periodic cube of length $L$ centered
at the origin. We initialize the domain as a uniform and stationary
fluid having rest mass density $\rho_0 = 1.0$ and gas pressure $p_g =
\rho_0/3$. We apply a uniform magnetic field along the $x$-direction
with magnitude $10^{-3}$. The flow is driven stochastically on large
scales according to a prescription we have adapted from
\cite{Schmidt:2009p2393}. The driving mechanism is intended to mimic
the effect of larger flow structures in which our domain is embedded,
and should thus be time-correlated with the turnover time of the
largest eddies, which is of order one light-crossing time of the domain,
$T_{lc}=L/c$. We achieve smooth time correlation by advancing the Fourier
modes $\tilde{\vct{a}}(\vct{k},t)$ of the driving field according to
an Ornstein-Uhlenbeck process \citep{Uhlenbeck:1930p4602}
$d\tilde{\vct{a}}(\vct{k},t)$, which consists of a restoring force
together with a complex-valued Gaussian-distributed random-walking
term, $d\tilde{\vct{W}} (\vct k, t)$:
\begin{eqnarray*}
  d\tilde{\vct a}(\vct k,t) =
  -\tilde{\vct a}(\vct k,t)
  \frac{dt}{T_{lc}} + \sqrt{\frac{P_{RMS} \sigma^2(\vct k)}{T_{lc}}}
  \mathfrak P(\vct k) \cdot d\tilde{\vct{W}} (\vct k, t)&
\end{eqnarray*}
The projection operator, 
\begin{equation}
  \mathfrak P_{ij}(\vct k) = \zeta \mathfrak P^\perp_{ij}(\vct k)
  + (1-\zeta)\mathfrak P^\parallel_{ij}(\vct k)
\end{equation}
is applied to the vector deviate $d\tilde{\vct{W}} (\vct k, t)$ in
order to select compressive and vortical driving modes separately,
according to the parameter $\zeta$. In this study, we use $\zeta=1$
which corresponds to a purely vortical driving field. For a detailed
study how $\zeta$ effects the turbulence statistics, see
\cite{Federrath:2010p4876}.

The acceleration field is applied to the 4-velocity of the flow, $u^\mu$ at
every time step, $\vct{u}(\vct{x},t) \mapsto \vct{u}(\vct{x},t) +
\vct{a}(\vct{x},t) dt/u^0$, where the spatial realization is obtained
by taking the real part of the Fourier mode superposition
\begin{equation}
  \label{eqn:trigseries}
  \vct{a}(\vct{x},t) = \Re \left\{ \sum_{0<|\vct{k}|<K_F}
    \tilde{\vct{a}}(\vct k,t) \exp(i \vct{k} \cdot \vct{x}) \right\}.
\end{equation}
The spectral profile, $\sigma^2(\vct{k}) \propto k^6 e^{-8k/k_1}$
\citep{Vestuto:2003p3456, Lemaster:2009p2711} is normalized to unity
over the driven wavenumbers. The length scale of maximum driving,
$\ell_1=2\pi/k_1$, and the cutoff $\ell_F=2\pi/K_F$ are chosen to be
$L/4$, and $L/2$ respectively. The small subset of driven wavenumbers
is chosen for reasons of efficiency, since the evaluation of Equation
\ref{eqn:trigseries} scales $\propto (2K_F+1)^3$ and is carried out
frequently. The driving mechanism which results from this prescription
is time correlated for $T_{lc}$ and has RMS power given by $P_{RMS}$,
which throughout this study has been set to $0.05$, delivering a
fractional power per light-crossing time, $\langle \dot
E_{tot}/E_{tot}\rangle T_{lc}$ of between $8\%$ and $10\%$ during the
steady-state period of the run. This relatively mild driving power
results in a trans-sonic turbulent flow which reaches a quasi-steady
state after roughly $5$ light-crossing times. We note that no
correlation is enforced between real and imaginary parts of
$d\tilde{\vct{W}}(\vct k, t)$, and thus the driving field is
statistically helicity-free. We refer to the model presented here as
SATS1 with the resolution appended such that e.g., SATS1-1024 refers
to the Super-Alf\'{e}nic trans-sonic model $1$ at resolution of
$1024^3$ zones.

\section{Results}
\label{sec:results}

\subsection{Startup transient and quasi-steady evolution}
\label{sec:startup}
Our model of RMHD turbulence begins with spatially uniform
conditions. Therefore, the early evolution of the model is
characterized by a startup transient. This transient can be roughly
partitioned into three stages, which have been shown in Figure
\ref{fig:time-hist}. The forth stage is quasi-steady evolution of the
model, and is characterized by very weak magnetic field growth and a
thermalization rate balanced by the driving power which causes gradual
increase of the internal energy. The results presented in \S
\ref{sec:results} are obtained from six snapshots taken during this
phase of the evolution.

About one correlation time of the driving field ($T_{lc}$) passes
before its RMS value is reached. During this stage, gentle driving
gradually increases the mean fluid velocity, along with the sonic and
Alfv\'{e}nic Mach numbers. For these Mach numbers, we report the
relativistic generalizations \citep{Gedalin:1993p2713}
\begin{eqnarray}
  \mathcal{M}_{s,A} &=& \frac{\beta_{f}(1-\beta_{f}^2)^{-1/2}}{\beta_{s,A}(1-\beta_{s,A}^2)^{-1/2}} \\
  \beta_s^2 &=& \frac{\Gamma p_g}{\rho h} \\
  \beta_A^2 &=& \frac{B^2}{B^2 + 4 \pi \rho h}
\end{eqnarray}
where $\beta_f$ is the fluid velocity.

This stage lasts for only about one $T_{lc}$, until the driving field
is fully ``warmed up.'' The second stage is characterized by
exponential growth of the magnetic energy, and linear growth of
kinetic energy. This means that the mean Alfv\'{e}n velocity increases
more rapidly than the bulk velocity, resulting in a sharp drop of the
Alfv\'{e}nic Mach number. The transition to quasi-steady evolution
occurs in the third stage and is complete at $4.7 T_{lc} =
T_{sat}$. During the quasi-steady phase, thermalization at small
scales balances the input of kinetic energy at large scales, causing
the pressure to gradually increase from $0.333$ to $0.720$ over the
remaining $\sim 7 T_{lc}$ duration of this quasi-steady
evolution. Also during this stage, magnetic structures gather
coherence over larger scales, as demonstrated in Figure
\ref{fig:pspec-ratio}(c). The fraction of magnetic to total (including
thermal) energy increases to $1.5\%$ throughout this phase, and the
equipartition scale ($P_K(k) \sim P_B(k)$) reaches $1/5$ the driving
scale. Evolution of a $512^3$ model (SATS1-512) through $24 T_{lc}$
indicates that the magnetic energy fraction stops growing before it
reaches $2.0\%$ at which time equipartition occurs at $1/3$ the
driving scale. Further discussion of this process is provided in \S
\ref{sec:pspec-mag}.

\begin{figure}
  \centering
  \includegraphics[width=\ScaleIfNeeded]{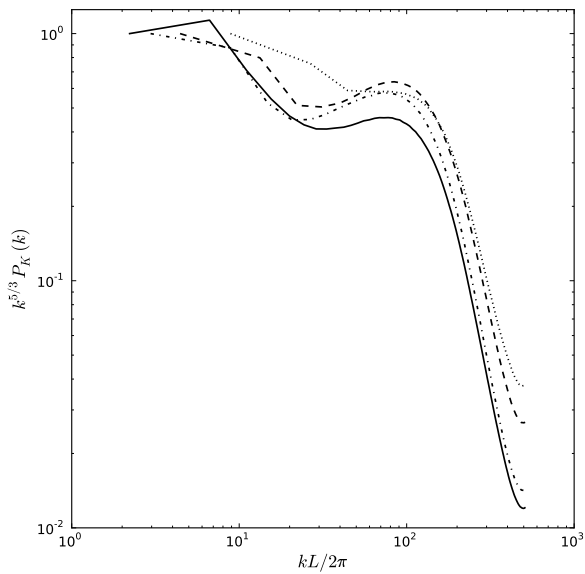}
  \caption{Convergence study for the power spectrum of the velocity,
    $P_K(k)$. Shown are the averaged power spectra, compensated by
    $k^{5/3}$ for the same model at three different resolutions,
    $512^3$ (\emph{dashed}), $768^3$ (\emph{dash-dotted}), and
    $1024^3$ (\emph{solid}).}
  \hspace{0.04\textwidth}
  \label{fig:pspec-converge-K}
\end{figure}

\subsection{Universality and locality of compressible turbulence}
Universal properties of a turbulent system are those which do not
dependent upon any boundary, initial, driving, or dissipative
conditions. The goal of many numerical turbulence investigations is to
establish universal relations within turbulent fields at different
length scales, since by extrapolation they can be used to characterize
natural systems whose Reynolds numbers are far greater than those
reached by the numerical model. Universal behavior is expected to
emerge when the outer and inner scales of turbulence are separated far
enough that effects of driving and dissipation are ``forgotten'' in an
intermediate range. Exactly what is meant by ``far enough'' depends on
the degree with which interactions can take place between eddies of
disparate sizes. When only eddies of comparable sizes are able to
exchange energy, the cascade is said to be highly localized. On the
other hand, if energy is capable of flowing directly between
structures separated by large wavenumber, the system is said to
exhibit nonlocality in its cascade. For such systems, enormously large
numerical resolution may be required in order to resolve universal
behavior.

The locality hypothesis for incompressible flows
\citep{Kolmogorov:1941p912} led to the well-verified prediction that
the power in the velocity field, $P_K(k)$ obeys a power law $\propto
k^{-5/3}$ throughout the inertial range. There are now strong
arguments that locality holds for compressible turbulence as well
\citep{Aluie:2011p4489}. In both compressible and incompressible MHD,
the dynamics of the cascade are substantially more complicated, and
nonlocality or ``diffuse locality'' \citep{Beresnyak:2009p4471,
  Beresnyak:2010p4470} is expected to hinder the emergence of
universal scalings in presently available numerical experiments. With
regard to the likelihood that our simulations are capable of resolving
universality, we suggest two reasons to be optimistic. The first is
that \cite{Porter:2002p4511} have shown that at $\mathcal{M}_S \sim 1$
the turbulent energy transfer is still consistent with the
\cite{Kolmogorov:1941p912} theory for incompressible turbulence,
meaning that locality is likely to be obeyed. The second is that the
nonlocality observed by \cite{Beresnyak:2009p4471} is for Alfv\'{e}nic
turbulence in the presence of a strong background magnetic field. For
our conditions of fully developed super-Alfv\'{e}nic turbulence with
no substantial mean field, \cite{Verma:2005p4519} predict that the
energy transfer is local as long as the net helicity is zero. Since
our driving mechanism is helicity-free, this assumption is certainly
met. Thus while power law scalings are not guaranteed, they should
also not be ruled out \emph{a priori} on the basis of distant
wavenumber interactions.
\begin{figure}
  \centering
  \includegraphics[width=\ScaleIfNeeded]{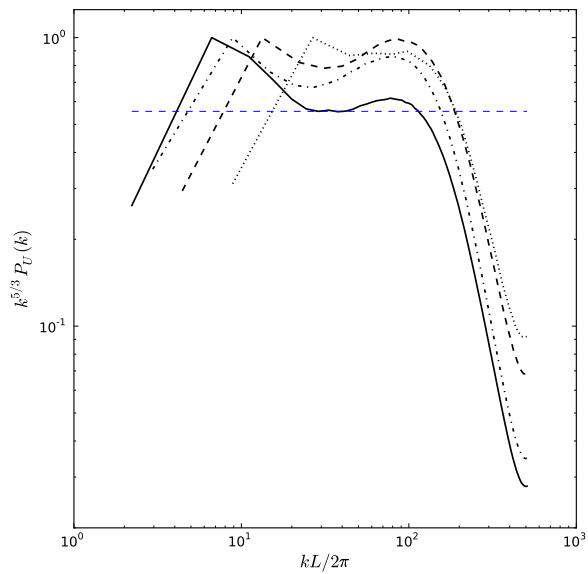}
  \caption{Convergence study for the power spectrum of the
    density-weighted velocity, $P_U(k)$ where $U = \rho^{1/3}v$. Shown
    are the averaged power spectra, compensated by $k^{5/3}$ for the
    same model at three different resolutions, $512^3$
    (\emph{dashed}), $768^3$ (\emph{dash-dotted}), and $1024^3$
    (\emph{solid}).}
  \hspace{0.04\textwidth}
  \label{fig:pspec-converge-U}
\end{figure}

\subsection{Convergence study for the power spectrum}
Here we assess the degree to which our simulations have resolved the
dynamics of the inertial interval by studying the power spectrum of
the same model at different grid resolutions. Careful judgement must
be used when searching for power law scalings in turbulence power
spectra. Generally, one searches the power spectrum for a ``flat''
interval between two systematic over-densities of power at both ends
of the spectrum. At high wave numbers, the so-called ``bottleneck''
effect \citep{Falkovich:1994p3874} causes a pile-up of power between
the inertial and dissipation scales. At low wave numbers, a sort of
inverse bottleneck occurs as energy rushes away from the driving
scale.

We have computed power spectra of the velocity, $P_K(k)$ and the
density-weighted velocity \citep{Kritsuk:2007p3858} $P_U(k)$, where $U
\equiv \rho^{1/3} v$ for the same model at the resolutions $N_{res} =
256^3$, $512^2$, $768^3$, and $1024^3$. It is important to point out
that our interpretation of the resolution study is that the size of
the grid spacing is kept fixed between runs, so that the box size is
reduced to $L \times N_{res}/1024$, and the driving field is moved to
higher wavenumber. In this interpretation, the power spectra are
compared to one another at the same number of light-crossing times of
the largest box size.

Figure \ref{fig:pspec-converge-K} shows the power in the velocity
field at each resolution. The main observation is that no inertial
range is uncovered for the velocity field, but that the bottleneck
(peaking at $kL/2\pi \sim 10^2$) is becoming less pronounced at higher
resolution. Figure \ref{fig:pspec-converge-U} on the other hand,
presents a compelling case that inertial behavior has been resolved
for the density weighted velocity $\rho^{1/3} v$ put forward by
\cite{Kritsuk:2007p3858}, and further studied in
\cite{Kowal:2007p4879, Schmidt:2008p4878, Federrath:2010p4876}. We
believe that this demonstrates consistency with the prediction of
their simple cascade model that $P_U(k) \propto k^{-5/3}$, as well as
their numerical findings for highly compressible hydrodynamical
turbulence. We note that \cite{Lemaster:2009p2711} have also measured
$P_U(k)$ in trans-Alfv\'{e}nic MHD turbulence at $\mathcal{M_S} \sim
6$ and obtained a much shallower slope of $-1.29$. We interpret this
as evidence that the cascade model of \cite{Kritsuk:2007p3858} may be
applicable to super-Alfv\'{e}nic, but not trans--Alfv\'{e}nic
flows. This interpretation is consistent with observations made by
\cite{Boldyrev:2002p4520} that the low-order hydrodynamic statistics
of super-Alfv\'{e}nic turbulence should closely resemble those from
the purely hydrodynamical case.

\subsection{Power in compressive versus vortical motions}
\begin{figure}
  \centering
  \includegraphics[width=\ScaleIfNeeded]{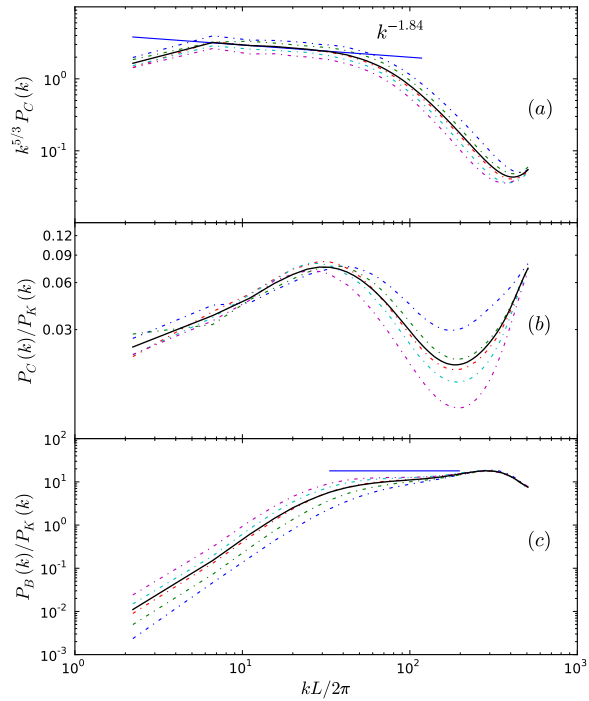}
  \caption{Different power spectra taken over the range of times
    $T_{lc}=8,9,10,11,12$ (\emph{dash-dotted}) and the corresponding
    time averaged power spectrum (\emph{solid}). (a): $P_C(k)$, the
    power in compressive velocity modes at wavenumber $k$, compensated
    by $k^{5/3}$. The blue line has a slope $-1.84$ and is fit over
    the wave numbers $k/2\pi \in [7,31]$. Later times are at lower
    value than earlier times, although the shape of the spectrum is
    not changing in time. (b): $P_C(k) / P_K(k)$, the ratio of
    compressive to total power in the velocity field. Snapshots taken
    at later times dip lower at moderate to high wavenumber. (c):
    $P_B(k) / P_K(k)$, the ratio of magnetic to kinetic power at
    wavenumber $k$. The trend is that this ratio increases in time at
    low and moderate wavenumber.}
  \hspace{0.04\textwidth}
  \label{fig:pspec-ratio}
\end{figure}
As the Mach number of turbulence increases, so does the degree of
fluid compressibility. This means that more of the power in the
velocity field is contained in compressive (dilating or shock-like)
structures. In order to observe the scale dependence of this effect,
we have decomposed the velocity field into solenoidal (curl-like) and
dilatational (divergence-like) parts using a Helmholtz decomposition,
\begin{subequations}\label{eqn:spec-decomp}
  \begin{align}
    P_C(\vct{k}) &= |\hat{\vct{k}} \cdot  \tilde{\vct{v}}_{\vct{k}}|^2\\
    P_S(\vct{k}) &= |\hat{\vct{k}} \times \tilde{\vct{v}}_{\vct{k}}|^2
  \end{align}
\end{subequations}
The power $P_C(k)$ in compressive modes is shown in Figure
\ref{fig:pspec-ratio}(a). $P_C(k)$ follows a power law $\propto
k^{-1.84}$ over the wavenumbers $k/2\pi \in [7,31]$, and no bottleneck
is observed. The lack of bottleneck in compressive modes has also been
observed by \cite[e.g.][]{Porter:1999p4875}, while a very similar
slope, $k^-{1.79}$, was observed in \cite{Federrath:2010p4876}. The
ratio $P_C(k) / P_K(k)$ of compressive to total power in the velocity
field is shown in Figure \ref{fig:pspec-ratio}(b). Due to the fact
that power is injected at large scales using strictly solenoidal
modes, there is an under-density of compressive power at low
wavenumber. As the details of energy injection are ``forgotten'' at
smaller scales, the power in compressive modes gradually increases
toward moderate wavenumber, reaching a maximum $9\%$ of the total
between the inertial and dissipative range. At yet higher wavenumbers
through the dissipative range, shearing motions become more
significant, but eventually give way to shocklets, causing another
rise in the compressive power near the grid scale. The overall trend
across scales is that $\sim 5\%$ of the total power is contained in
dilatational motion of the fluid, similar to what was found by
\cite{Porter:2002p4511} for $\mathcal{M}_S \sim 1$ flows. It is
interesting to note for non-relativistic highly supersonic
($\mathcal{M}_S \sim 10$) but otherwise very similar conditions,
\cite{Boldyrev:2002p4520} found that $P_C(k) / P_K(k) \sim 10\% -
20\%$ through the inertial range, observing a very similar profile to
that shown in Figure
\ref{fig:pspec-ratio}(b). \cite{Federrath:2010p4876} have studied the
effect of driving solenoidal versus dilatational modes in
$\mathcal{M}_S \sim 5$ models the ratios $1/3$ and $1/2$
respectively. In their study, these ratios are constant over the
inertial range.

As far as convergence to a quasi-stationary state is concerned, we
consider the curves in Figure \ref{fig:pspec-ratio}(a) to be robust,
because only the overall normalization is changing from one snapshot
to the next. However, there is a definite time trend in the shape of
$P_C(k) / P_K(k)$, indicating that further time evolution would be
required to obtain a time-converged measurement of the compressive to
total power ratio at different wavenumbers.

\subsection{Power spectrum of magnetic energy} \label{sec:pspec-mag}
Figure \ref{fig:pspec-ratio}(c) shows the ratio $P_B(k) / P_K(k)$ of
power in the magnetic field to power in the velocity field. The ratio
gradually increases from $\sim 1\%$ at the driving scale, through to
super-equipartition at the beginning of the dissipative range. It then
transitions to become constant across scales throughout the
dissipative range. Although the ratio is still changing in time, with
more coherence of magnetic structures occurring at large scale, we
have observed time converged behavior in a $512^3$ simulation which
was run over a longer time. In doing so, we conclude that the
interpretation of Figure \ref{fig:pspec-ratio}(c) is robust. After
saturation, the ratio of magnetic to kinetic power stays constant over
the dissipative range, and obeys a very straight power law throughout
the driving and inertial ranges, increasing from low to moderate
wavenumber. The scale at which the ratio $P_B(k) / P_K(k) = 1$ occurs
at $k \sim 5 K_F$ (see \S \ref{sec:initmodel-driving}) in the most
recent snapshot. However, the trend is for the equipartition scale to
move to lower wave number as magnetic structures form coherence over
larger scales. \cite{Cho:2000p4549} have observed that once fully
steady state is reached, the equipartition scale occurs at $k \sim 3
K_F$.

\subsection{One dimensional structure functions of the velocity}
\begin{figure}
  \includegraphics[width=\ScaleIfNeeded]{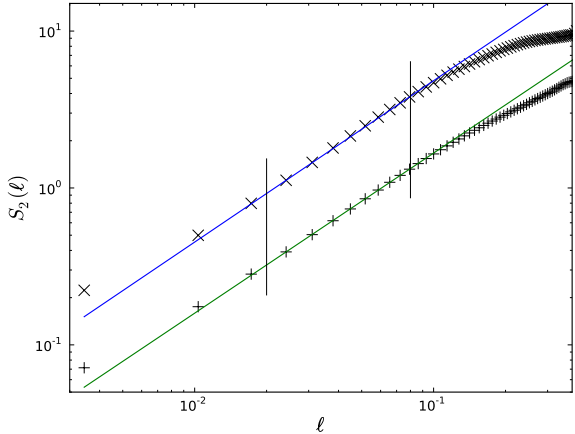}
  \caption{Shown is the one-dimensional, second order structure
    function of the velocity field $S_2^\perp(\ell)$ (\emph{$\times$'s})
    and $S_2^\parallel(\ell)$ (\emph{$+$'s}). They are fit by power
    laws $S_2(\ell) \propto \ell^{\zeta_2}$ with
    $\zeta_2^{\perp}=1.03$ and $\zeta_2^{\parallel}=1.02$, where
    $21\Delta < \ell < 82\Delta$.}
  \hspace{0.04\textwidth}
  \label{fig:struct1d}
\end{figure}
\begin{figure}
  \includegraphics[width=\ScaleIfNeeded]{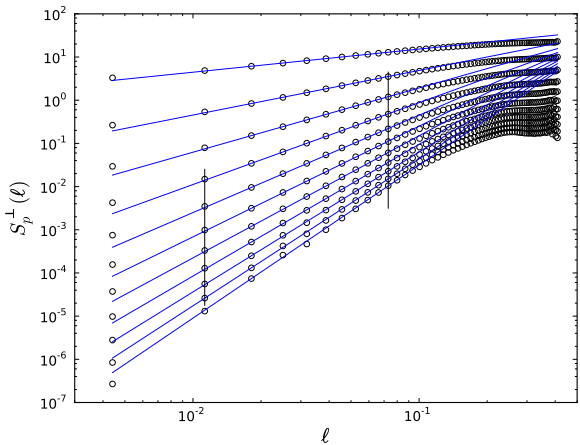}
  \caption{One-dimensional, $p$th order structure function of the
    velocity field $S_p^\perp(\ell)$, with $p=1$ (\emph{top}) to
    $p=11$ (\emph{bottom}).}
  \hspace{0.04\textwidth}
  \label{fig:she-leveque-Spl}
\end{figure}
\begin{figure}
  \includegraphics[width=\ScaleIfNeeded]{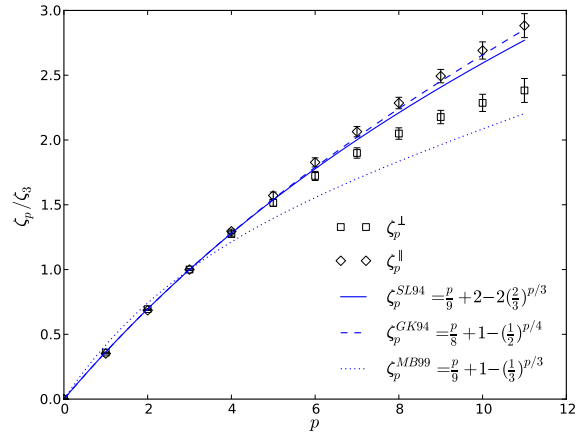}
  \caption{The slope $\zeta^\perp_p$ (\emph{boxes}) and
    $\zeta^\parallel_p$ (\emph{diamonds}) for different values of the
    order $p$. The error bars are obtained by shifting the fit window
    by a single bin. Shown also are the predictions of
    \cite{She:1994p4537} (\emph{solid}) for incompressible
    non-relativistic hydro turbulence, \cite{Grauer:1994p4541}
    (\emph{dashed}) and \cite{Muller:2000p4539} (\emph{dotted}) for
    incompressible MHD turbulence. All data have been normalized by
    $\zeta_3$.}
  \hspace{0.04\textwidth}
  \label{fig:she-leveque-zetap}
\end{figure}

We have measured the structure functions of the velocity field to
order $p$,
\begin{equation*}\label{eqn:padoan-strc}
  S_p^{\parallel, \perp}(\boldsymbol{\ell}) = \langle
  |v^{\parallel, \perp}(\mathbf{r}+\boldsymbol{\ell}) -
  v^{\parallel, \perp}(\boldsymbol{r})|^p \rangle
\end{equation*}
where the velocity vector is decomposed in parallel and perpendicular
components relative to the displacement vector
$\boldsymbol{\ell}$. \cite{She:1994p4537}, have determined based on
very general assumptions that these functions should scale $\propto
\ell^{\zeta_p}$ within the inertial range of fully developed
incompressible turbulence, where
\begin{equation}
  \zeta_p = \frac{p}{9} + 2 - 2(\frac{2}{3})^{p/3}
\end{equation}
These predictions have been extended analytically to incompressible
MHD turbulence \citep{Grauer:1994p4541} and verified numerically
\citep{Politano:1995p4542, Muller:2000p4539}. The compressible MHD
case has been studied numerically by \cite{Boldyrev:2002p3882} and
\cite{Padoan:2004p4405} in the context of supersonic molecular cloud
turbulence, finding remarkably good agreement with analytical
predictions.

Figure \ref{fig:struct1d} shows both $S_2^{\parallel}(\ell)$ and
$S_2^{\perp}(\ell)$. We find that over the range $21\Delta < \ell <
82\Delta$, the structure functions are described by a power law with
index $\zeta_2 = 1.025 \pm 0.005$. The high precision of this fit does
not rule out systematic errors, but it does imply that some degree of
scale-invariant behavior has been resolved in our simulations. The
range over which the fit is valid occurs at smaller scales than what
was reported by \cite{Kritsuk:2007p3858} ($32\Delta$ to $256\Delta$)
but very similar to those found by \cite{Boldyrev:2002p3882}
($10\Delta$ to $90\Delta$). The slope $\zeta_2 = 1.025$ is slightly
steeper than what was observed by \cite{Kritsuk:2007p3858}, and
considerably steeper than the value of $2/3$ which follows from the
\cite{Kolmogorov:1941p912} theory. We also observe that the ratio
$S_2^\perp / S_2^\parallel = 2.5$ is larger by a factor of $2$ than
what was seen in \cite{Kritsuk:2007p3858}, whose result is in closer
agreement with the \citep{Kolmogorov:1941p912} theory. This
observation warrants farther attention, since if the discrepancy is
not numerical in nature, but a feature of relativistic MHD turbulence
then it may provide important hints in seeking a relativistic
extension to the \cite{She:1994p4537} model.

Using about $10^9$ sample pairs, we have been able to compute the
slopes of higher order structure functions through $p=11$. Figure
\ref{fig:she-leveque-Spl} demonstrates the quality of these fits for
each order. We have used the same window to compute the slope of the
higher order structure functions as for the $S_2$ case. Figure
\ref{fig:she-leveque-zetap} shows the slopes $\zeta_p$ for the
transverse and longitudinal structure functions for each $p$,
normalized by $\zeta_3$. Also shown are the predictions of
\cite{She:1994p4537} for incompressible non-relativistic hydro
turbulence, \cite{Grauer:1994p4541} and \cite{Muller:2000p4539} for
incompressible MHD turbulence. We find a remarkable agreement between
the \cite{Grauer:1994p4541} prediction and the longitudinal velocity
fluctuations, even at the highest order. The data for transverse
velocity fluctuations lie midway between the \cite{Muller:2000p4539}
prediction.

\subsection{Scale-dependent anisotropy of the velocity field}

\begin{figure}
  \includegraphics[width=\ScaleIfNeeded]{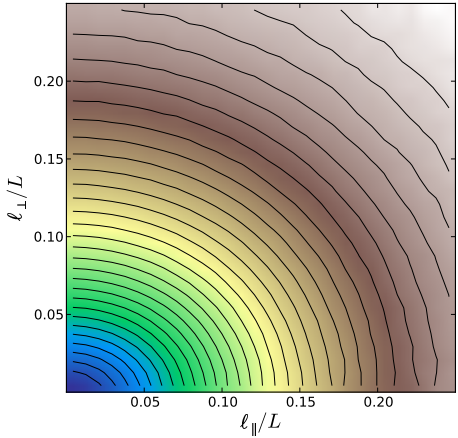}
  \caption{Shown is the two-dimensional, second order structure
    function of the velocity field $S^v_2(\ell_\perp,
    \ell_\parallel)$. The offset vector $\boldsymbol{\ell}$ is
    decomposed into components parallel \emph{($x$-axis)} and
    perpendicular \emph{($y$-axis)} to the local magnetic field,
    $\frac{1}{2}(\vsp{B}\left(\vsp{r}+\boldsymbol{\ell}) +
      \vsp{B}(\vsp{r})\right)$.}
  \hspace{0.04\textwidth}
  \label{fig:struct2d}
\end{figure}
MHD turbulence in the presence of a strong background field, known as
Alfv\'{e}nic turbulence, consists of an energy cascade mediated by
interacting MHD waves. This situation is treatable with the formalism
of wave turbulence \cite[see e.g.][]{EvgenevichZakharov:1992p4576},
whereby the resonant interactions between the MHD ``free particles''
(solutions to the linearized equations) are treated
perturbatively. For incompressible MHD, the only free particles are
the shear and pseudo Alfv\'{e}n waves. Analysis of the resonant
nonlinear interaction of these modes lead to the model of
\cite{Goldreich:1995p4506} which predicts a Kolmogorov-like energy
spectrum $\propto k^{-5/3}$. Their findings also included the
so-called scale-dependent anisotropy with respect to the local mean
field, $k_\parallel \propto k_\perp^{2/3}$. This phenomenon is
understood geometrically as the exaggerated distortion of eddies,
becoming more apparent at smaller scales. The numerical verification
was first obtained by \cite{Cho:2000p4507}, who presented two methods
of numerical measurement corresponding to the correct geometrical
interpretation. Other numerical studies have observed the same scaling
in supersonic MHD turbulence \citep{Cho:2003p3040,
  Beresnyak:2005p4444} and force-free relativistic Alfv\'{e}nic MHD
turbulence \citep{Cho:2005p4453}, the latter having received prior
analytical treatment by \cite{Thompson:1998p4548}. Here we provide the
corresponding measurement for super-Alfv\'{e}nic, compressible
relativistic MHD turbulence.

We use the same method proposed by \cite{Cho:2000p4507} to measure the
scale-dependence of eddy distortions, which relies on the
eccentricities of level-surfaces for the second order structure
functions
\begin{eqnarray}
  S^v_2(\ell_\perp, \ell_\parallel) &=& \langle |\vsp{v}(\vsp{r}+\boldsymbol{\ell}) -
  \vsp{v}(\vsp{r})|^2 \rangle \\
  S^B_2(\ell_\perp, \ell_\parallel) &=& \langle |\vsp{B}(\vsp{r}+\boldsymbol{\ell}) -
  \vsp{B}(\vsp{r})|^2 \rangle
\end{eqnarray}
where $\boldsymbol{\ell}$ is decomposed into cylindrical coordinates
oriented along the local mean field, which is defined for each pair of
points as $\frac{1}{2}(\vsp{B}\left(\vsp{r}+\boldsymbol{\ell}) +
  \vsp{B}(\vsp{r})\right)$.

In examining the shape of eddies, \cite{Cho:2000p4507} used the
structure functions for the velocity and magnetic field $S^v_2$ and
$S^B_2$, finding slopes which were on average slightly larger than
$2/3$ for $S^v_2$ and slightly smaller for $S^B_2$. As depicted in
Figure \ref{fig:magP}, coherent magnetic field structures exist out to
only about $1/10$ of the domain size, meaning that $S^B_2$ provides a
rather narrow window over which to measure the scaling. However, the
velocity field is coherent out the driving scale which affords a broad
range of scales over which $S^v_2$ scales robustly.

\begin{figure}
  \includegraphics[width=\ScaleIfNeeded]{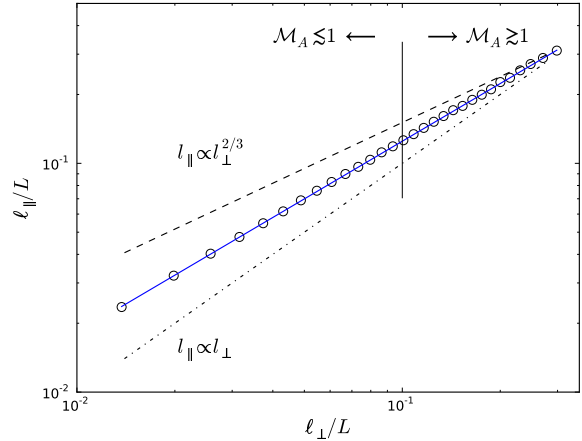}
  \caption{The semi-major ($\ell_\parallel$) and semi-minor
    ($\ell_\perp$) axes of the elliptical contours obtained from
    Figure \ref{fig:struct2d} (\emph{circles}), and the best fit line
    (\emph{solid}) $\ell_\parallel \propto \ell_\perp^{0.84}$. For
    comparison we provide the prediction of \cite{Goldreich:1995p4506}
    $\ell_\parallel \propto \ell_\perp^{2/3}$ (\emph{dashed}) and a
    slope of unity (\emph{dash-dotted}). The vertical line marks the
    scale $\ell = L/10$ at which the magnetic and kinetic energy are
    in equipartition.}
  \hspace{0.04\textwidth}
  \label{fig:scale-aniso}
\end{figure}

Indeed, we observe excellent scaling in the structure functions of
velocity, indicating that the slope $\ell_\parallel \propto
\ell_\perp^{0.84}$ is very robust. In fact, the validity of the fit is
valid \emph{at least} between $12\Delta$ and $300\Delta$, the limiting
factor in extending the scaling being due to data collection
techniques. The fact that the slope is steeper than the $2/3$ means
that the eddy distortions depend upon the scale more weakly than in
the \cite{Goldreich:1995p4506} model. This may suggest corrections to
the cascade dynamics to account for relativistic effects, but we
forgo any conclusions on this matter until a detailed comparative
study with the equivalent non-relativistic case has been completed.

It is also very interesting that the same power law slope of $0.84$
holds above and also below the equipartition scale. Under
non-relativistic super-Alfv\'{e}nic conditions, \cite{Cho:2000p4549,
  Cho:2003p3040} both find similar contour shapes to those shown in
Figure \ref{fig:struct2d}. However, in those studies the contour
intercepts were not provided, meaning that the precise scaling
behavior above and below the equipartition scale is uncertain to
us. \cite{Beresnyak:2005p4444} does provide these scalings for
trans-Alfv\'{e}nic, supersonic conditions, but only below the
equipartition scale. We believe that a rigorous study of scaling above
and below the equipartition scale in super-Alfv\'{e}nic turbulence is
required in order to isolate the effects of compressibility,
substantial magnetic field curvature, and also relativistic effects.

\section{Conclusions} \label{sec:conclusions}

We have measured spectral and scaling properties of relativistically
warm magnetohydrodynamic turbulence in the mildly compressible and
super-Alfv\'{e}nic regime. The numerical models were simulated at very
high resolution ($1024^3$) using Mara, a new second order Godunov code
tuned for accurate and robust evolution of the RMHD equations in three
dimensions. Our main production model was driven stochastically at
large scales during quasi-steady evolution for about $6$
light-crossing times of the domain. We find that:

\begin{enumerate}
\item The magnetic energy is amplified from seed fields to $1.5\%$ of
  the total fluid energy. The scale at which equipartition between
  magnetic and kinetic energy occurs is between $1/5$ and $1/3$ of the
  driving scale.
\item At $1024^3$ the power spectrum of velocity is dominated by a
  bottleneck, but not inconsistent with the Kolmogorov prediction of
  $k^{-5/3}$. The power spectrum of density-weighted velocity
  $\rho^{1/3} v$ scales $\propto k^{-5/3}$ over moderate wavenumbers,
  consistent with the simple cascade model of
  \cite{Kritsuk:2007p3858}.
\item About $5\%$ of kinetic energy is in compressive modes. These
  modes follow a power law over large to moderate scales with index
  $-1.84$.
\item The transverse and longitudinal one dimensional structure
  functions of velocity are well fit by a power law over moderate to
  small scales. As a function of the order $p$, the slope of
  longitudinal velocity fluctuation is well described by the
  prediction of \cite{She:1994p4537}. Statistically significant
  deviation is observed for the transverse fluctuation.
\item Mild elongation of coherent velocity structures along the local
  magnetic field is observed. The degree of elongation is
  scale-dependent, but more weakly than is predicted by
  \cite{Goldreich:1995p4506}. The scale dependence obeys a power law
  to high precision above and below the equipartition scale.
\end{enumerate}

These results suggest that for trans-sonic, super-Alfv\'{e}nic
relativistic astrophysical conditions the turbulent cascade dynamics
share many similarities with their non-relativistic
counterparts. However, the high order scaling relations developed for
non-relativistic media, as well as the Alfv\'{e}nic cascade model of
\cite{Goldreich:1995p4506} may require modification in order to be
applicable to the relativistic astrophysical environments.  A detailed
comparison between non-relativistic and relativistic MHD models is
currently in progress and will form the basis for a future
publication.

\acknowledgments

This research was supported in part by the NSF through grant
AST-1009863 and by NASA through grant NNX10AF62G issued through the
Astrophysics Theory Program. Resources supporting this work were
provided by the NASA High-End Computing (HEC) Program through the NASA
Advanced Supercomputing (NAS) Division at Ames Research Center. We
would like to thank the Institute for Theory and Computation at the
Harvard-Smithsonian Center for Astrophysics for hospitality, where a
portion this work was completed. We thank Andrey Beresnyak for helpful
suggestions regarding measurement techniques, and Paul Duffell for
many useful discussions. We would also like to acknowledge the
developers of the Python $\mathtt{numpy}$, $\mathtt{matplotlib}$, and
$\mathtt{h5py}$ modules, which were used extensively in our figures
and analysis.

\bibliographystyle{apj}
\bibliography{papers}

\end{document}